\documentclass[usenatbib,usegraphicx]{mn2e}

\title[The progenitor set of present-day early-type galaxies]
    {The progenitor set of present-day early-type galaxies}

\author[S.~Kaviraj et al.]
{S. Kaviraj$^{1}$\thanks{E-mail: skaviraj@astro.ox.ac.uk},
  J. E. G. Devriendt$^{2}$, I. Ferreras$^{3}$, S. K. Yi$^{4}$ and J. Silk$^{1}$\\
$^{1}$Department of Physics, University of Oxford, Keble Road,
  Oxford OX1 3RH, UK\\
$^{2}$Observatoire Astronomique de Lyon, 9 Avenue Charles Andr\'e,
  69561 Saint-Genis Laval cedex, France\\
$^{3}$Physics Department, King's College London, Strand, London,
WC2R 2LS\\
$^{4}$Center for Space Astrophysics, Yonsei University, 134
Shinchon, Seoul 120-749, Korea}

\begin{document}

\date{11 January 2006}

\pagerange{\pageref{firstpage}--\pageref{lastpage}} \pubyear{2003}

\maketitle

\label{firstpage}


\begin{abstract}
We present a comprehensive theoretical study, within a fully
realistic semi-analytical framework, of the photometric properties
of early-type progenitors in the redshift range $0<z<1$, as a
function of the luminosity and local environment of the early-type
remnant at present-day. We find that while larger early-types are
generally assembled \emph{later}, their luminosity-weighted
stellar ages are typically \emph{older}. In dense environments,
$\sim70$ percent of early-type systems are in place by $z=1$ and
evolve without major interactions thereafter, while in the field
the corresponding value is $\sim30$ percent. Averaging across all
environments at $z\sim1$, less than 50 percent of the stellar mass
which ends up in early-types today is actually in early-type
progenitors at this redshift. The corresponding value is $\sim65$
percent in clusters due to faster morphological transformations in
the such dense environments. We also develop probabilistic
prescriptions which provide a means of \emph{including} spiral
(i.e. non early-type) progenitors at intermediate and high
redshifts, based on their luminosity and optical ($BVK$) colours.
For example, at intermediate redshifts ($z\sim0.5$), large
($M_B<-21.5$), red ($B-V>0.7$) spirals have $\sim75-95$ percent
chance of being a progenitor, while the corresponding probability
for large blue spirals ($M_B<-21.5$, $B-V<0.7$) is $\sim50-75$
percent. Finally, we explore the correspondence between the
\emph{true} progenitor set of present-day early-types and the
commonly used `red-sequence', defined as the set of galaxies
within the part of the colour-magnitude space which is dominated
by early-type objects. While large members ($M_V<-22$) of the `red
sequence' trace the progenitor set accurately in terms of numbers
and mass, the relationship breaks down severely at fainter
luminosities ($M_V>-21$). Hence the red sequence is generally
\emph{not} a good proxy for the progenitor set of early-type
galaxies.
\end{abstract}




\section{Introduction}
As `end points' of galaxy merger sequences, early-type galaxies
carry important signatures of mass assembly and star formation in
the Universe. Deducing their star formation histories (SFHs)
therefore contains the key to understanding not only the evolution
of these galaxies but characteristics of galaxy formation as a
whole. Our view of early-type galaxy formation has developed over
the years, away from the classical version of `monolithic
collapse' and towards the hierarchical assembly of these objects
through mergers and accretion of fragments over a Hubble time, as
advocated by the currently popular LCDM paradigm of galaxy
formation.

Traditionally, the evolution in the properties of early-type
galaxies, for example their optical colours, has been traced by
studying early-type populations at progessively higher redshift.
However, a fundamental feature of early-type formation in the
standard model is that stellar mass that eventually ends up in
present-day early-type galaxies is not \emph{entirely} contained
in early-type systems at high redshift. Looking \emph{only} at
early-type systems at high redshift introduces a \emph{progenitor
bias}, which becomes increasingly more severe at higher redshift,
as the fraction of early-type progenitors becomes progressively
smaller \citep{VD2001,VD1996}.

In the current era of large scale (optical) surveys e.g. SDSS,
COMBO-17, GALEX, MUSYC, GEMS, unprecedented amounts of data
spanning a large range in redshift and environment is becoming
available, allowing us to study statistically significant numbers
of galaxies at various stages of evolution. A quantitative
assessment of progenitor bias is therefore needed to gauge the
role of non-early-type progenitors, especially for studies which
centre on high redshift.

A central theme of this work is to quantify progenitor bias. This
phenomenon has already been studied by \citet{VD2001}. However,
their study employed phenomenological SFHs, with the simple
assumption that morphological transformations occur $\sim$ 1.5
Gyrs after the cessation of star formation in a particular galaxy.
Our study extends the results of \citet{VD2001} by using a fully
realistic semi-analytical framework, in which mass assembly and
morphological transformation can be followed more accurately in
the context of the currently popular $\Lambda$CDM paradigm. We
study the evolution of the \emph{progenitor set} (galaxies that
are progenitors of present-day early-types) with redshift, as a
function of the luminosity and environment of the early-type
remnant which is left at present-day. We pay particular attention
to \emph{spiral progenitors} in the model, since these are
routinely excluded from studies of early-type galaxies at high
redshift by virtue of their morphology, even though they form an
important part of the progenitor set. By comparing the properties
(optical colours and luminosities) of spiral progenitors to the
general spiral population, we provide a means of correcting for
progenitor bias by \emph{including} specific parts of the spiral
population at high redshift into the study of early-type
evolution.

The plan of this paper is as follows. In Section 2, we quantify
the morphology of the progenitor set and map the general
properties of elliptical, S0 and spiral progenitors as a function
of redshift. In Section 3 we focus exclusively on \emph{spiral
progenitors} and compare their photometric properties to the
general spiral population. In Section 4 we trace the contribution
of galaxies in dense regions (groups and clusters) at high
redshift to cluster early-types at present-day. In Section 5 we
explore the correspondence between the \emph{true} progenitor set
of present-day early-type galaxies and the `red-sequence', defined
as the set of galaxies within the part of the colour-magnitude
space dominated by early-type objects, which is commonly
used as a proxy for the progenitor set \citep[e.g.][]{Bell2004}.

Note that throughout this study we provide \emph{rest-frame}
magnitudes for all model galaxies. Unless otherwise noted, the
filters used are in the standard Johnson system.


\section{Dissecting the progenitor set: morphologies of progenitors}
Early-type galaxies have an assortment of SFHs. Crucial to this
study is the \emph{dynamical age} of an early-type galaxy, which
we define as the epoch at which its last merger took place. This
last merger creates the early-type remnant and imparts its final
early-type morphology. As our subsequent analysis of progenitor
bias involves morphologies and environments of model galaxies, a
brief explanation of the definition of these quantities is
necessary.

Galaxy morphology in the model is determined by the ratio of the
B-band luminosities of the disc and bulge components which
correlates well with Hubble type \citep{Simien1986}. A morphology
index is defined as

\begin{equation}
I = \exp\bigg(\frac{-L_B}{L_D}\bigg)
\end{equation}

such that a pure disc has $I=1$ and a pure bulge has $I=0$.
Following \citet{Baugh1996}, ellipticals have $I<0.219$, S0s have
$0.219<I<0.507$ and spirals have $I>0.507$. This simple
prescription is clearly incapable of capturing the complex
spectrum of real galaxy morphologies. Therefore, in what follows,
`spirals' refer to \emph{all} systems which do not have a dominant
spheroidal (bulge) component. Observationally, this includes not
only systems with distinctive spiral morphologies, but also
peculiar or irregular systems.

Galaxy environments in the model are driven by the mass of the
dark matter (DM) halo in which they are embedded. At $z=0$, DM
halo masses greater than $\sim 10^{14} M_{\odot}$ correspond to
`cluster' environments, while halo masses between $\sim 10^{13}
M_{\odot}$ and $\sim 10^{14} M_{\odot}$ correspond to `groups'.
All other halo masses correspond to the `field'. At higher
redshifts these definitions do not strictly hold since the DM halo
population is evolving - for example, the largest haloes at $z=1$
are likely to be roughly half their size at present day
\citep[e.g][]{VandenBosch2002}. We take this mass accretion
history into account when specifying the environments of galaxies
at high redshift. The mass accretion history is taken from
\citet[][see their Figure 5]{VandenBosch2002}.

\begin{figure}
\begin{center}
$\begin{array}{c}
\includegraphics[width=3.5in]{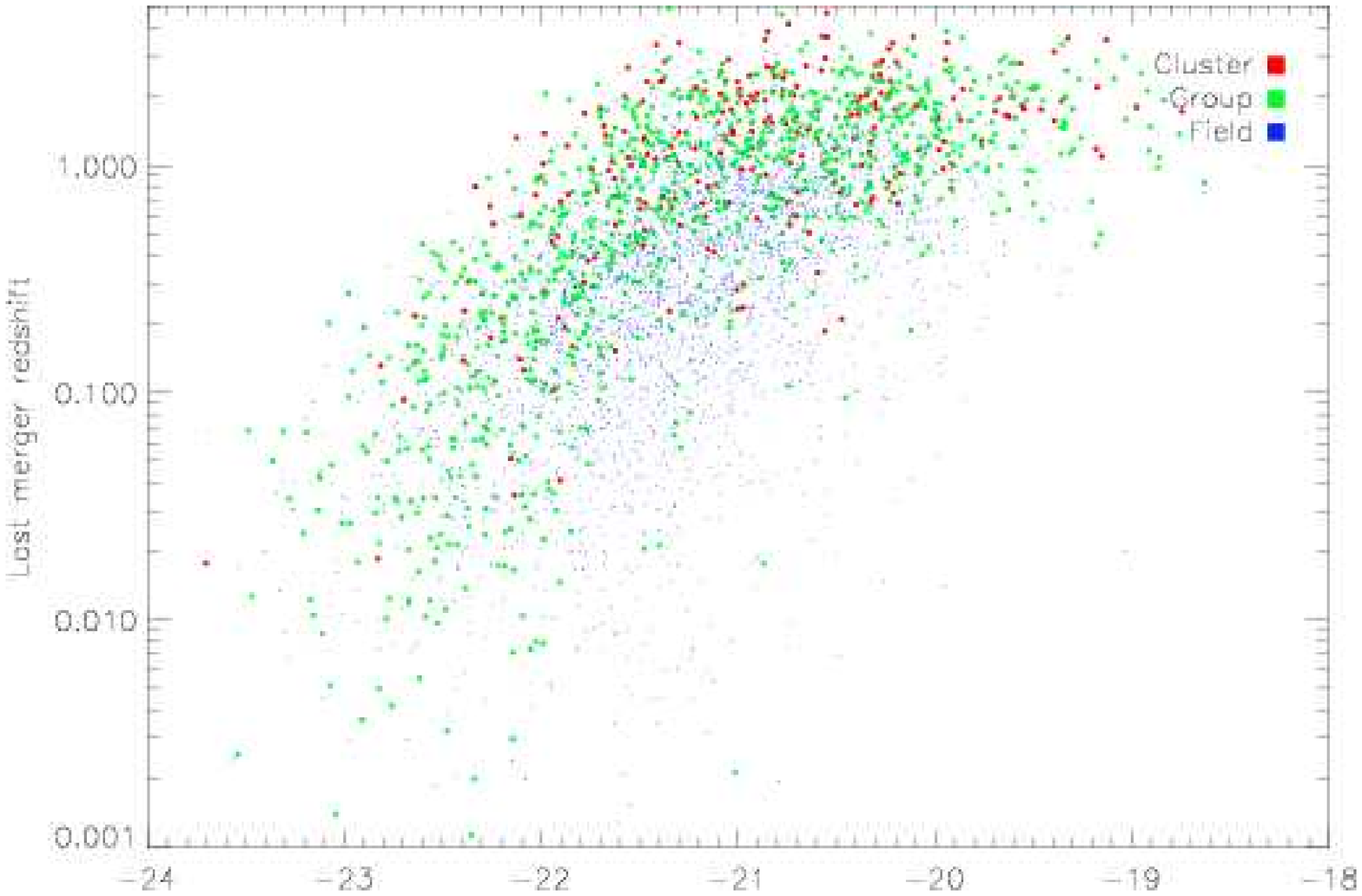}\\
\includegraphics[width=3.5in]{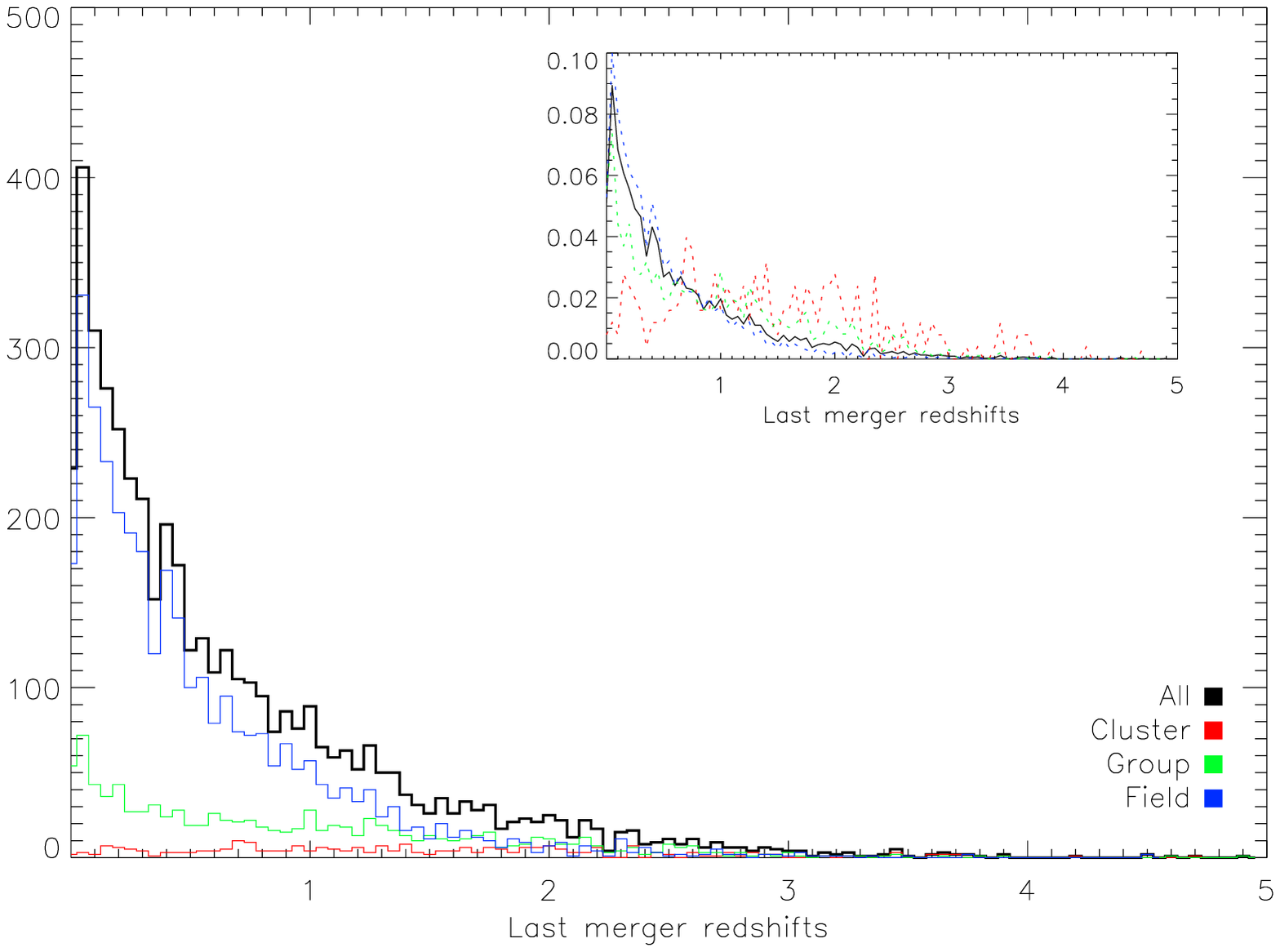}
\end{array}$
\caption{TOP: Last merger redshifts of the sample of early-types
in the model split by environment of the remnant at $z=0$. BOTTOM:
Histograms of last merger redshifts shown in the top panel, split
by environment. The inset shows the \emph{fraction} of mergers at
a given redshift.}
\label{fig:last_mergers}
\end{center}
\end{figure}

\begin{figure}
\begin{center}
\includegraphics[width=3.5in]{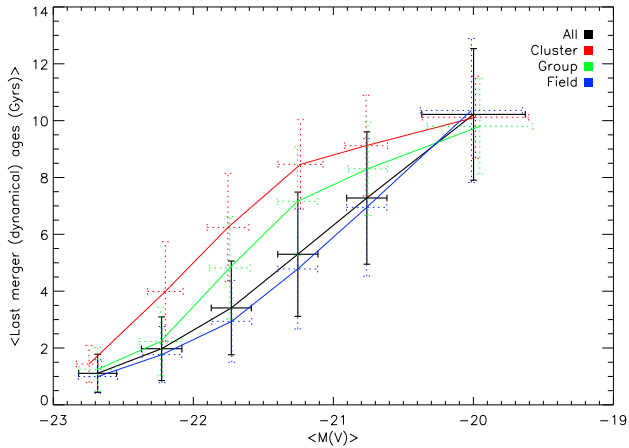}
\caption{Last merger ages as a function of average luminosity and
environment of the early-type remnant at $z=0$.}
\label{fig:av_lmages}
\end{center}
\end{figure}

\begin{figure}
\begin{center}
\includegraphics[width=3.5in]{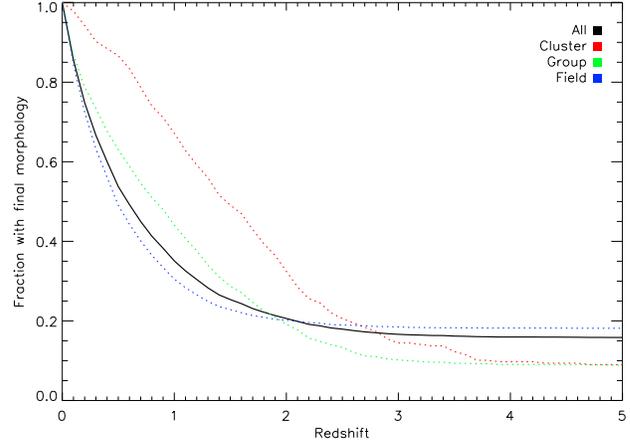}
\caption{Cumulative fraction of early-type galaxies which have has
their last merger as a function of redshift.}
\label{fig:final_morphs}
\end{center}
\end{figure}

Figure \ref{fig:last_mergers} (top panel) indicates the last
merger redshifts of the sample of early-types in the model split
by \emph{environment of the remnant} at $z=0$. The bottom panel in
Figure \ref{fig:last_mergers} shows histograms of the last merger
redshifts shown in the top panel, again split by environment. The
inset shows the \emph{fraction} of mergers at a given redshift. In
Figure \ref{fig:av_lmages} we plot the \emph{average} last merger
ages as a function of the luminosity and environment of the
early-type remnant at $z=0$. As expected, we find that, in all
environments, larger early-types are \emph{assembled later}
(although their stars are generally older \citep{Kaviraj2005a}).
Cluster galaxies (at least those brighter than $L_*$) have
significantly larger dynamical ages - morphological
transformations in clusters proceed more quickly than in all other
environments. This point is more clearly made in Figure
\ref{fig:final_morphs}, where we plot the \emph{cumulative}
fraction of early-type galaxies which have already has their last
merger. These galaxies have therefore achieved their final
`early-type' morphology and are evolving `monolithically'. We find
that, on average, without reference to environment, only 35
percent of early-type galaxies are in place by $z=1$ (black line)
- the rest are still `in pieces'. In terms of morphological
transformations cluster environments are \emph{special}, in that
early-type morphologies are attained significantly faster in
clusters (red curve), with almost 70 percent of early-type
galaxies having undergone their last merger by $z=1$.

\begin{figure}
\begin{center}
\includegraphics[width=3.5in]{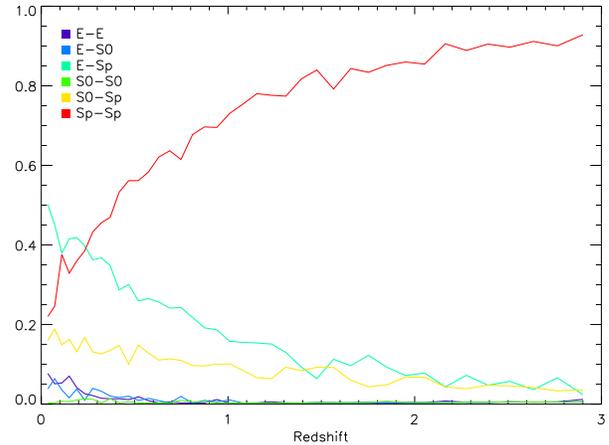}
\caption{Morphologies of progenitors in binary mergers as a
function of redshift. Non-binary mergers do happen but are rare,
and only take place at redshifts greater than $z\sim1.5$. Mergers
at intermediate and high redshift are dominated by pairs of
progenitors which contain at least one spiral progenitor.}
\label{fig:merger_types}
\end{center}
\end{figure}

Before the last merger occurs, the morphology of the progenitors
is \emph{not necessarily early-type}. Figure
\ref{fig:merger_types} shows the morphologies of progenitors in
\emph{binary} mergers as a function of redshift. Non-binary
mergers do happen but are rare, and only take place at redshifts
greater than $z\sim1$. In the local Universe, mergers between
early-type progenitors make up less than 20 percent of the merger
activity. All other mergers contain at least one spiral
progenitor. Mergers involving solely spiral progenitors
increasingly dominate at higher redshift and dominate the merger
activity beyond $z=1$.

\begin{figure}
\begin{center}
\includegraphics[width=3.5in]{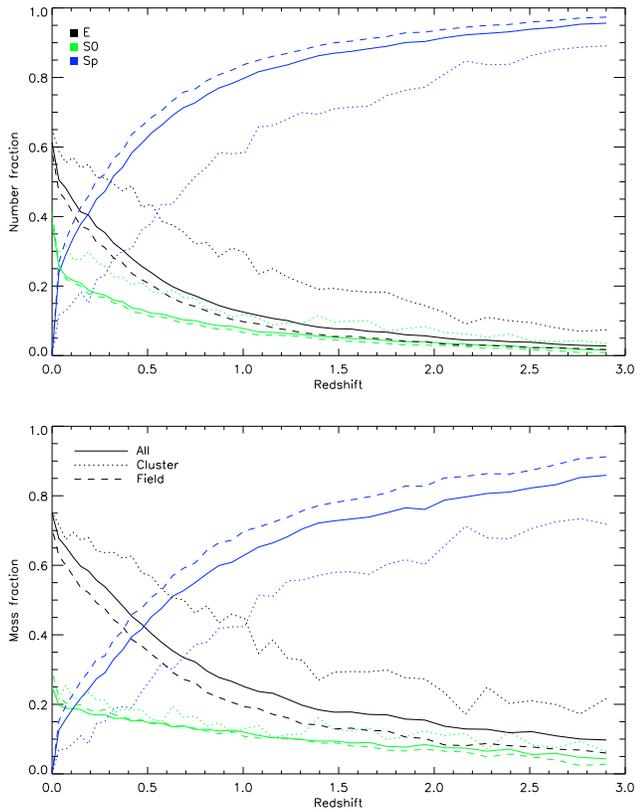}
\caption{Number (top) and mass (bottom) fractions contained in
progenitors of different morphological types in the redshift range
$0<z<3$, split by environment of the early-type remnant.}
\label{fig:num_mass_env}
\end{center}
\end{figure}

\begin{figure}
\begin{center}
\includegraphics[width=3.5in]{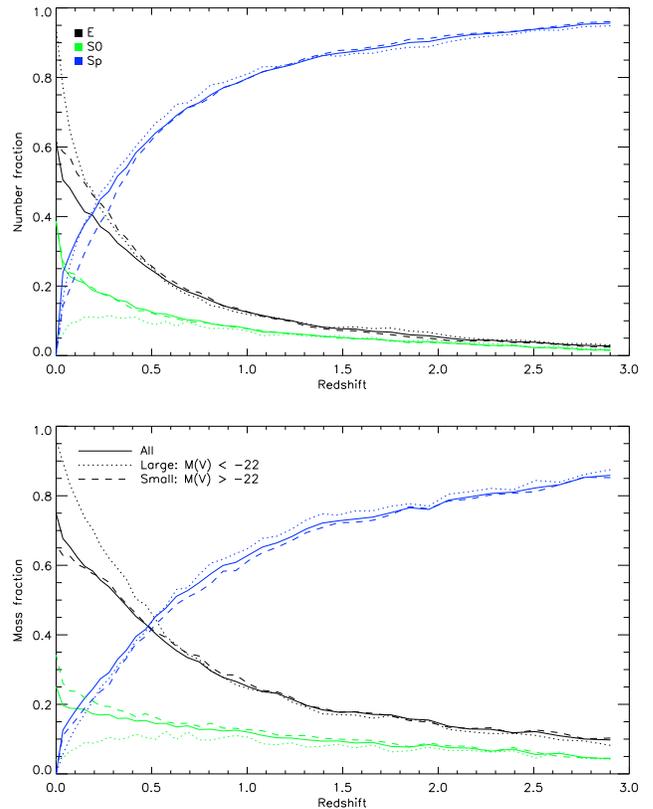}
\caption{Number (top) and mass (bottom) fractions contained in
progenitors of different morphological types in the redshift range
$0<z<3$, split by luminosity of the early-type remnant.}
\label{fig:num_mass_lum}
\end{center}
\end{figure}

Having provided a picture of the merger activity within the
progenitor set, it is instructive to look at the \emph{fraction}
of the progenitor set which is made up of a certain morphological
type as a function of redshift. Figures \ref{fig:num_mass_env} and
\ref{fig:num_mass_lum} shows the number and mass fractions
contained in progenitors of different morphological types in the
redshift range $0<z<3$, split by environment and luminosity of the
early-type remnant at $z=0$ respectively. We find that, averaging
across all environments, at $z\sim 1$, less than 50 percent of the
stellar mass which ends up in early-types today is actually in
\emph{early-type progenitors} at this redshift. Faster
morphological transformation in cluster environments means that
this value is $\sim65$ percent in clusters at $z\sim1$. As a
result, looking \emph{only} at early-type galaxies at $z\sim1$
does not take into account almost half the stellar mass in the
progenitor set. In other words, the mass in the progenitor set
doubles between $z=1$ and $z=0$. A similar result was found by
Bell at al. (2004), although they used the optical `red sequence'
in their study, which does not completely correspond to the
progenitor set of present-day early-type galaxies (see section 5
below).

\begin{figure}
\begin{center}
\includegraphics[width=3.5in]{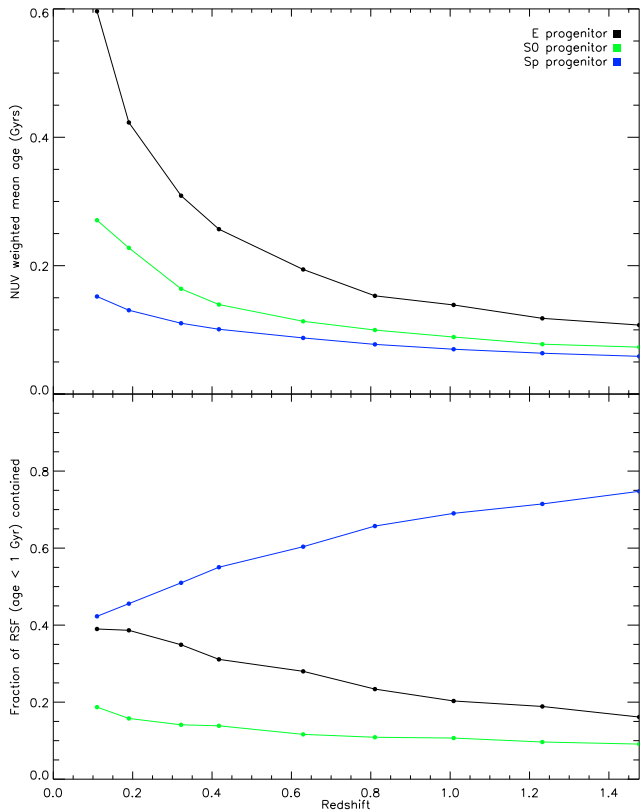}
\caption{TOP: Average $NUV$-weighted ages of progenitors of
different morphological types. The $NUV$ weighting is heavily
dominated by stars formed in these progenitors within the last 0.5
to 1 Gyr of look-back time. Note that the $NUV$ weighting was
generated using the GALEX (Martin el al. 2005) $NUV$ filter.
BOTTOM: The fraction of recently formed stars (age $<1$ Gyrs old)
across the progenitor set which is contained in progenitors of
each morphological type.} \label{fig:rsf_contained}
\end{center}
\end{figure}

The bias does not arise simply because some progenitor mass is not
taken into account, but because the age profile of the mass in
progenitors of different morphological types tends to vary. We
illustrate this point in Figure \ref{fig:rsf_contained}. The top
panel shows the average $NUV$-weighted ages of progenitors of
different morphological types. The $NUV$ weighting, generated
using the GALEX (Martin el al. 2005) $NUV$ filter, is heavily
dominated by stars formed within the last 0.5 to 1 Gyr of
look-back time. At all redshifts, early-type progenitors have
higher $NUV$-weighted ages, because the mass fraction contributed
by recent star formation (RSF) i.e. within the last 1 Gyr is
smaller than for spiral progenitors. The differences between
elliptical and spiral progenitors are most pronounced at low
redshift. The bottom panel shows the fraction of the RSF across
the progenitor that is contained in each morphological type. This
plot obviously has to be interpreted in conjunction with the mass
fractions hosted by each morphological type as a function of
redshift. For example, at $z \sim 0.1$, although spiral
progenitors host $\sim$ 40 percent of the total RSF in the
progenitor set, they only constitute $\sim$ 30 percent of the mass
in the progenitor set (see bottom panel of Figure
\ref{fig:num_mass_env}. Early-type progenitors (elliptical and S0
taken together) contribute $\sim60$ percent of the RSF - but they
also constitute $\sim 70$ percent of the total mass in the
progenitor set. Therefore at $z\sim0.1$ spiral progenitors host
1.5 times the amount of RSF \emph{per unit mass} than their
early-type counterparts. At higher redshift the balance of RSF
contained in each morphological type moves towards spiral
progenitors, partly because they are more spirals in the Universe
than early-types.

Figure \ref{fig:rsf_contained} illustrates that an increasingly
larger fraction of RSF in the progenitor set is contained in
late-type systems at increasing redshift. In the context of
colour-magnitude relations (CMRs), which are often used to
age-date early-type populations at all redshifts, the exclusion of
spiral progenitors at high redshift biases the CMR towards redder
colours and does not give a proper indication of the age of the
stellar mass that eventually constitutes present-day early-type
galaxies.

\section{The spiral progenitors}
One of the aims of this study is to provide a means of
\emph{including} spiral populations observed at high redshift
which might be early-type progenitors, and thus correct, at least
partially, for progenitor bias. We therefore focus on spiral
progenitors predicted by the model and compare their photometric
properties to the general spiral population.

\begin{figure}
\begin{center}
\includegraphics[width=3.5in]{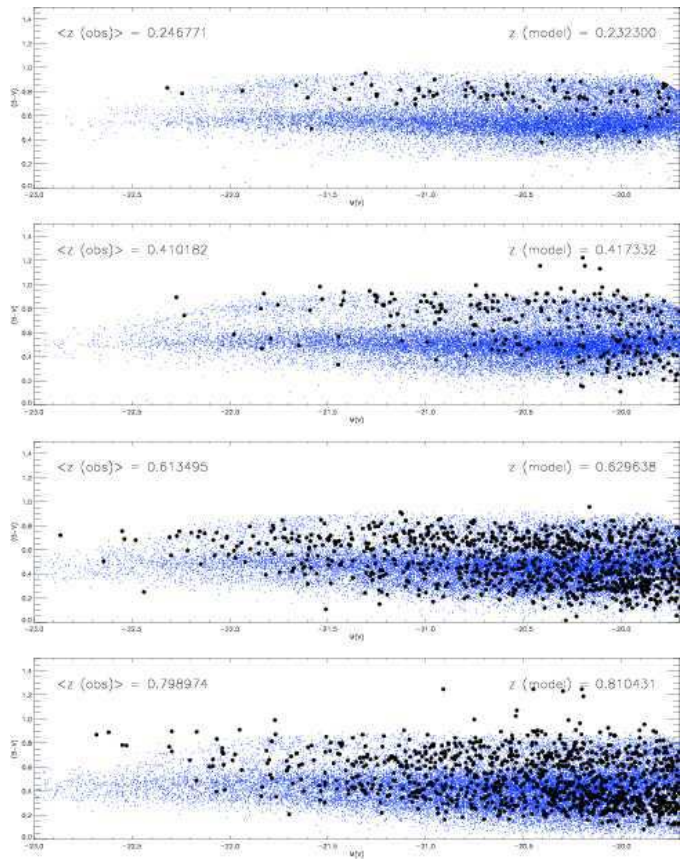}
\caption{Comparison between the colours of COMBO-17 galaxies and
the predicted spiral population in GALICS. The COMBO-17 sample is
restricted to galaxies which correspond to the GALICS completeness
limits in the B and V bands of -18.9 and -19.7 mags respectively.}
\label{fig:sam_combo_compare}
\end{center}
\end{figure}

Providing reasonably accurate prescriptions for finding
progenitors (of any morphology) based on predicted photometry is
clearly reliant on the predictions being reasonably representative
of the observed data. The correspondence between predictions from
GALICS and observed optical photometry (mainly at low redshift) is
shown in \citet[][Section 8]{Hatton2003}. The predictions from
GALICS produces good agreement to the galaxy luminosity functions
observed by the 2dF survey in the B and K bands \citep{Cross2001}.
The $(B-V)$ colours of spiral galaxies closely follow the observed
data of \citet{Buta1994}, both in terms of average values and
scatter. The version of the model used in this study is the same
as that used in \citet{Kaviraj2005a}, who calibrated GALICS to
accurately reproduce the optical colours of elliptical galaxies in
dense environments from $z=0$ to $z \sim 1.23$. Since our study
hinges on photometric predictions at high redshift, we compare, as
a further check, in Figure \ref{fig:sam_combo_compare}, the
$(B-V)$ colours of the spiral population in GALICS to recent
photometry from COMBO-17 survey \citep{Wolf2004} in the redshift
range $0<z<1$. The COMBO-17 sample used is restricted to galaxies
which correspond to the GALICS completeness limits in the B and V
bands of -18.9 and -19.7 mag respectively.

We find that predicted spiral colours in GALICS correspond well to
the range of observed $(B-V)$ colours in the COMBO-17 survey in
the redshift range $0<z<0.8$. Comparison to higher redshifts is
not possible because the accuracy of the observed V magnitudes
cannot be guaranteed beyond $z\sim0.7$ \citep{Wolf2004}. However,
these comparisons give us some confidence that the colour
parameter space spanned by the predicted spiral population in
GALICS is consistent with observations for a wide range of
redshifts.

\subsection{The luminosity function of spiral progenitors}
\begin{figure*}
\begin{center}
\begin{minipage}{126mm}
\includegraphics[width=\textwidth]{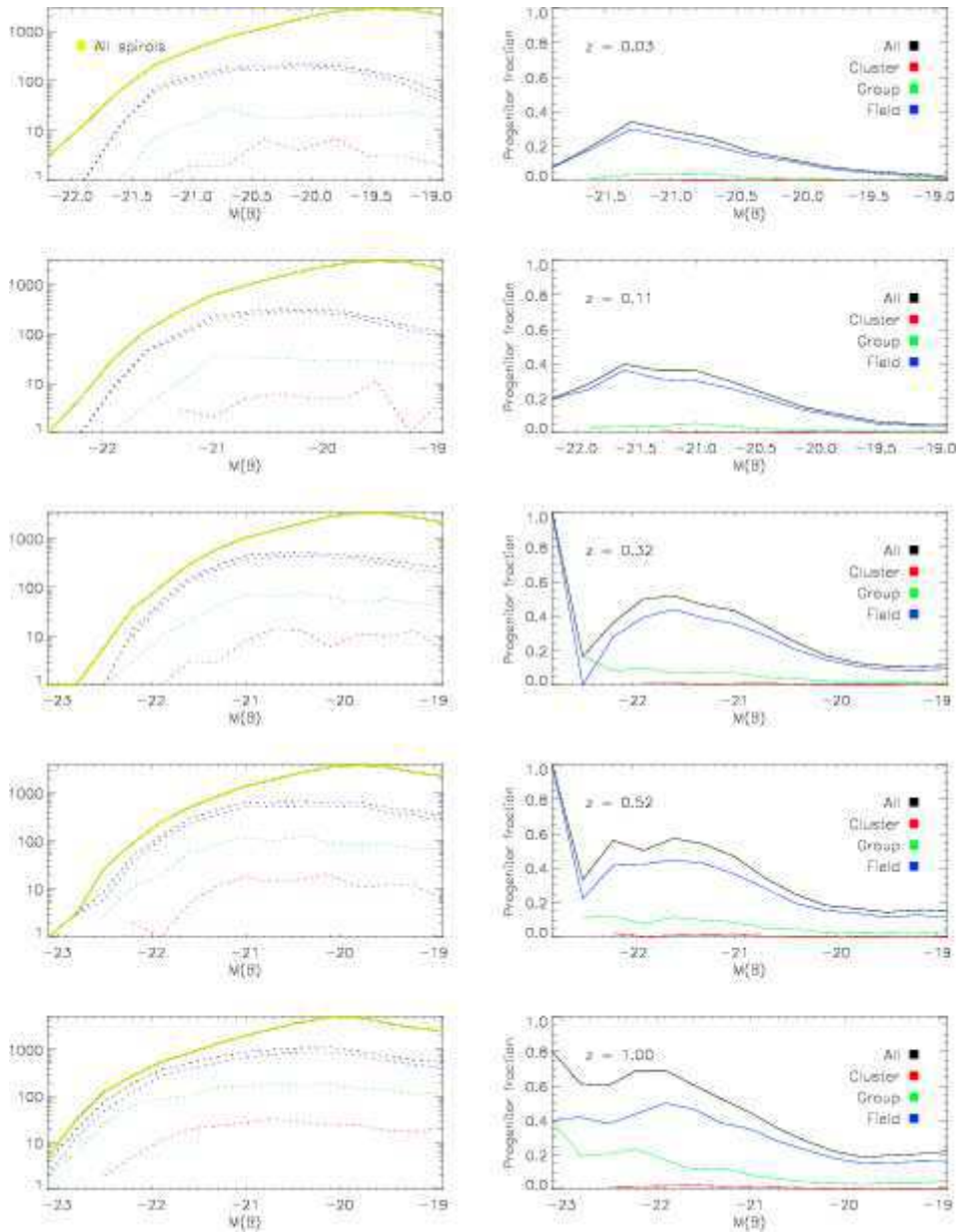}
\caption{The B-band luminosity function of spiral progenitors.}
\label{fig:b_spiralprogs}
\end{minipage}
\end{center}
\end{figure*}

\begin{figure*}
\begin{center}
\begin{minipage}{126mm}
\includegraphics[width=\textwidth]{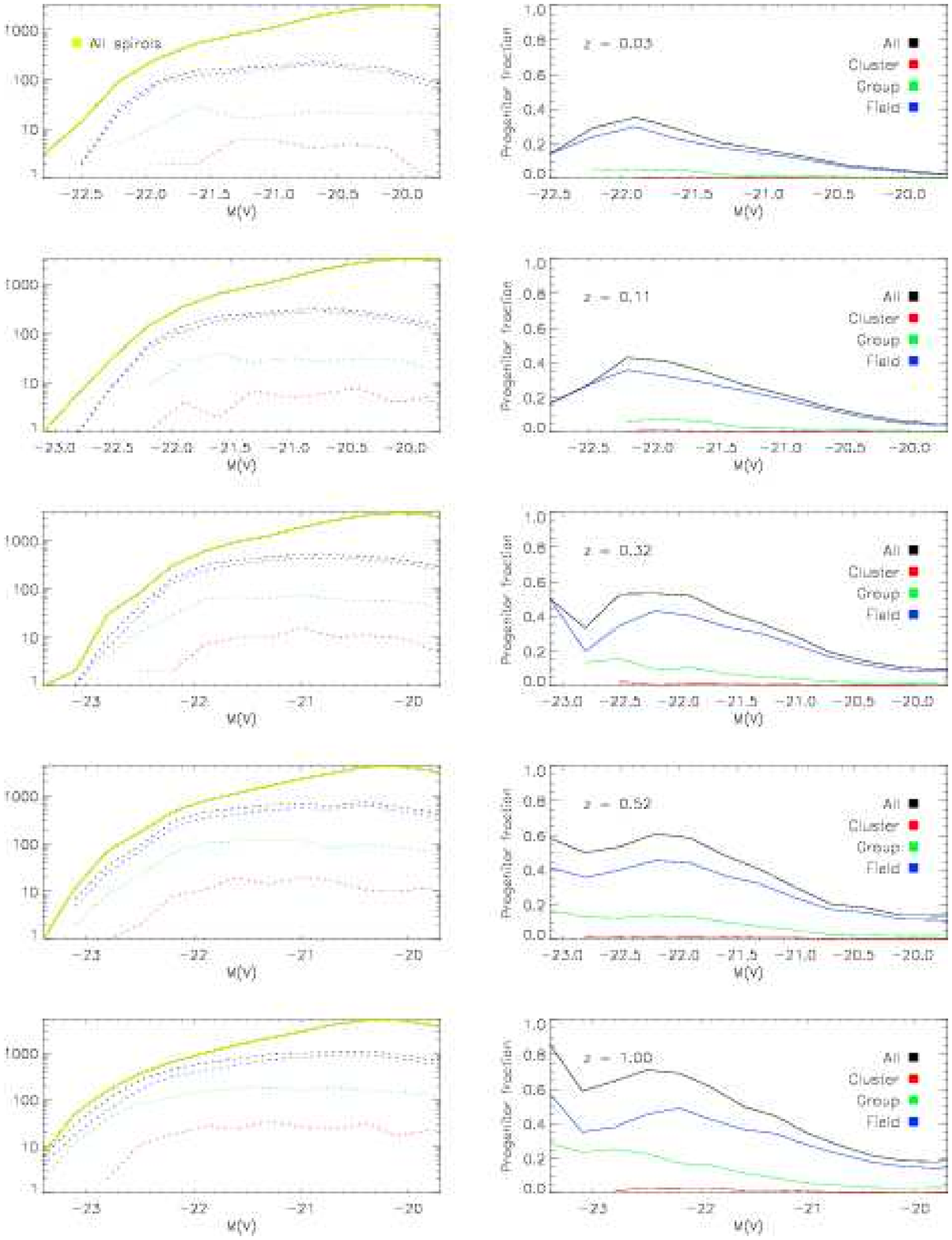}
\caption{Same as Figure \ref{fig:v_spiralprogs} but for the
$V$-band.} \label{fig:v_spiralprogs}
\end{minipage}
\end{center}
\end{figure*}

\begin{figure*}
\begin{center}
\begin{minipage}{126mm}
\includegraphics[width=\textwidth]{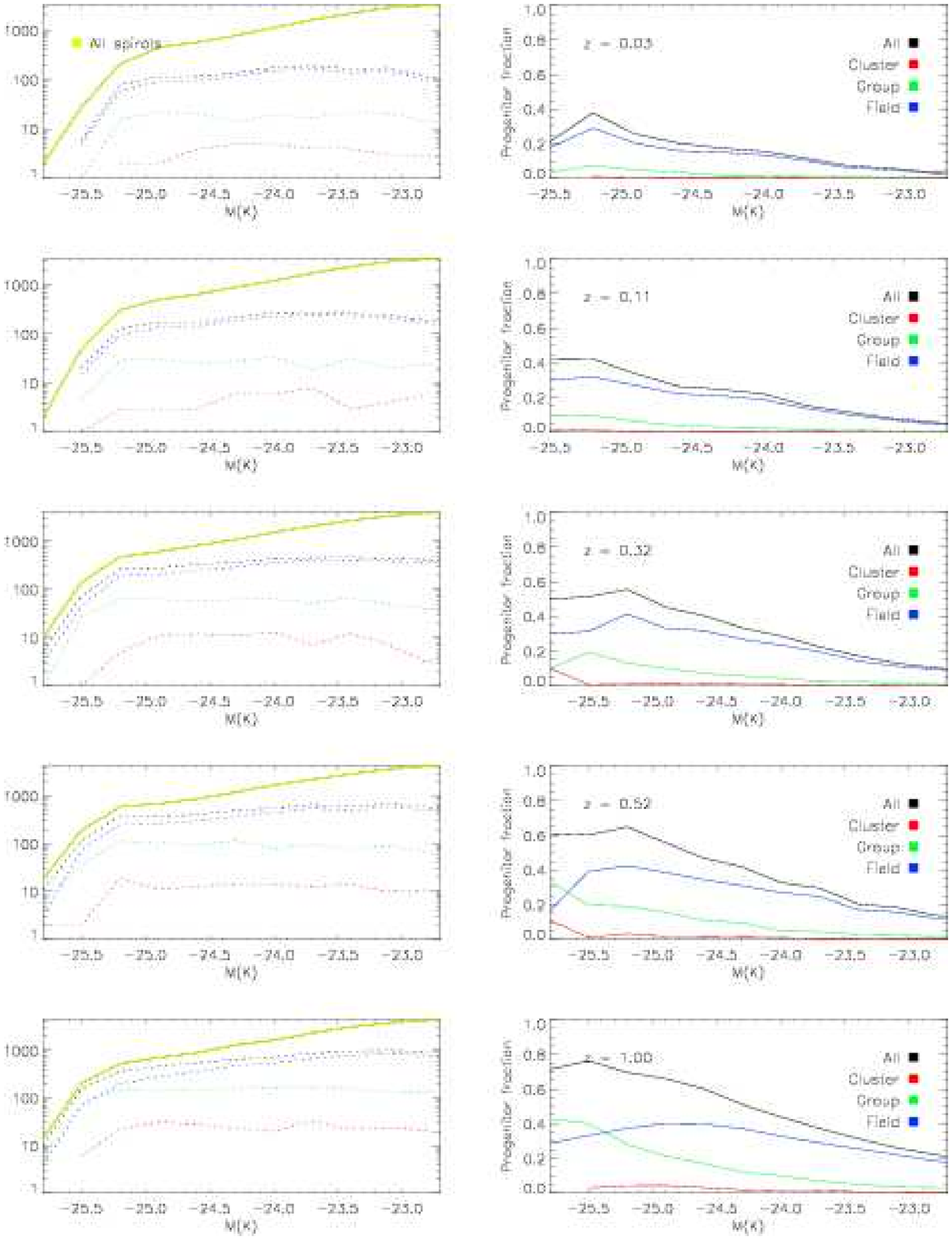}
\caption{Same as Figure \ref{fig:k_spiralprogs} but for the
$K$-band.}
\label{fig:k_spiralprogs}
\end{minipage}
\end{center}
\end{figure*}

We begin by studying the luminosity function (LF) of spiral
progenitors. We are interested in studying how the luminosities of
spiral progenitors compare to the general spiral population and
what fraction of spirals, at a given luminosity, are early-type
progenitors. In Figure \ref{fig:b_spiralprogs} we show the
evolution of the $B$-band LF of spiral progenitors - we also show
the spiral progenitors separated by the environment of their
present-day early-type remnant. The left hand column illustrates
the evolution of the spiral LFs - the yellow curve denotes the LF
of the general spiral population and the black curve the LF of all
spiral progenitors. The LFs of spiral progenitors whose early-type
remnants are, at $z=0$, in clusters, groups and the field, are
shown in red, green and blue respectively. Figures
\ref{fig:v_spiralprogs} and \ref{fig:k_spiralprogs} show the
corresponding plots for the predicted $V$-band and $K$-band
photometry respectively.

It is apparent that there is a greater preponderance of
progenitors among larger spirals at all redshifts. For example, at
low redshifts ($z<0.l$), 20 to 40 percent of spirals with
$M(B)<-20.5$ are early-type progenitors. At intermediate redshifts
($0.3<z<0.52$), these values rise to 30 and 60 percent
respectively. At high redshift ($z\sim1$) spirals with
$M(B)<-21.5$ have more than a 60 percent probability of being an
early-type progenitor, while spirals with $-20<M(B)<-21.5$ have
between a 30 and 40 percent chance of being early-type
progenitors. The falling progenitor fractions towards lower
redshift are partly due to the changing morphological mix of the
Universe.


\subsection{The colour magnitude space of spiral progenitors}
While investigating the LFs of spiral progenitors is useful in
indicating the \emph{probability} that a spiral of a given
luminosity has of being a progenitor, it is also desirable to
explore the colour-magnitude (CM) space of the spiral population,
so that we can separate progenitor spirals better from the general
population at a given luminosity.

\begin{figure*}
\begin{center}
\begin{minipage}{126mm}
$\begin{array}{c}
\includegraphics[width=\textwidth]{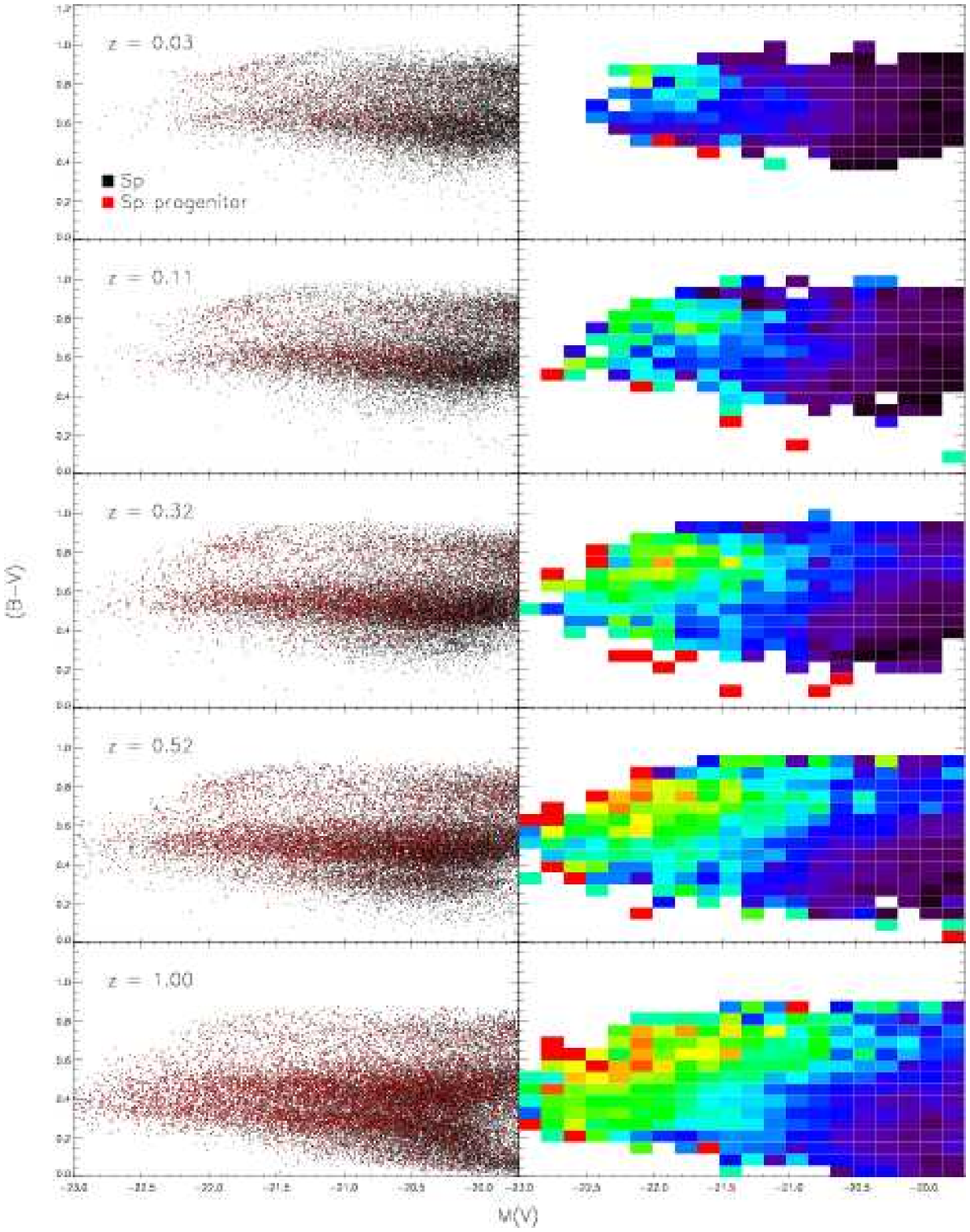}
\includegraphics[width=0.2\textwidth]{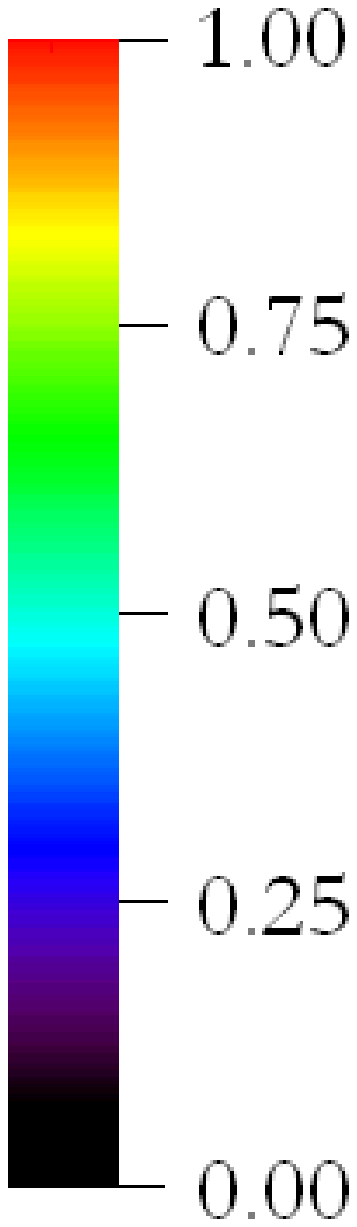}
\end{array}$
\caption{$(B-V)$ colours of the general population compared to the
$(B-V)$ colours of those spirals which are progenitors of
present-day early-type galaxies. Black dots represent spiral
galaxies and red dots represent spiral progenitors. The left-hand
column shows the spiral $(B-V)$ CMR from $z=0$ to $z=1$. In the
right hand column we show the fraction of spirals in parts of the
$(B-V)$ CMR space which are progenitors of early-type galaxies.
The fractions are indicated using the colour coding shown in the
legend. Warmer colours indicate a higher progenitor fraction (red
implies a progenitor fraction of 1, black represents a progenitor
fraction of 0 and parts of the CM space without any galaxies are
not colour-coded).} \label{fig:bv_spiralprogs}
\end{minipage}
\end{center}
\end{figure*}

In Figure \ref{fig:bv_spiralprogs} we compare the $(B-V)$ colours
of the general spiral population to the $(B-V)$ colours of spiral
progenitors. The left-hand column shows the spiral $(B-V)$ CMR
from $z=0$ to $z=1$. Black dots represent the spiral galaxies and
red dots represent spiral progenitors. In the right hand column we
show the fraction of spiral progenitors across the $(B-V)$ CMR.
The fractions are indicated using the colour coding shown in the
legend. Warmer colours indicate a higher progenitor fraction (red
implies a progenitor fraction of 1, black represents a progenitor
fraction of 0 and parts of the CM space without any galaxies are
not colour-coded).

At local redshifts ($z\sim0.03$), spirals brighter than
$M(B)\sim-21$ have $\sim30$ percent chance of being an early-type
progenitor, irrespective of their $(B-V)$ colour. At $z\sim0.1$,
spirals with $-21.5<M(B)<-20.5$ have $\sim30$ percent chance of
being a progenitor. For larger spirals, those with red $(B-V)$
colours, i.e. $(B-V)>0.8$, have $\sim60$ percent chance of being a
progenitor, while the corresponding probability for bluer spirals
is 30 to 50 percent.

\begin{figure*}
\begin{center}
\begin{minipage}{126mm}
$\begin{array}{c}
\includegraphics[width=\textwidth]{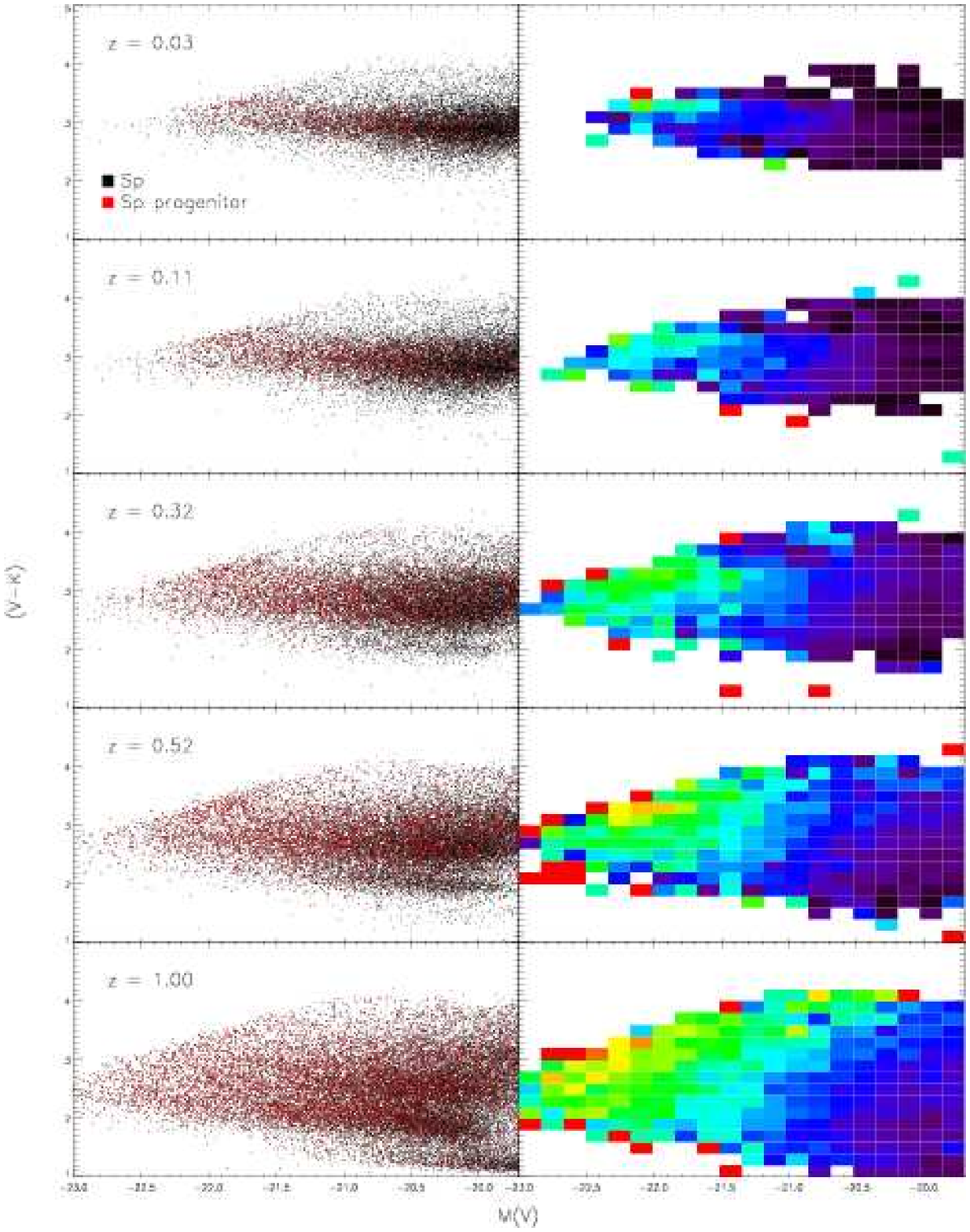}
\includegraphics[width=0.2\textwidth]{progfrac_legend}
\end{array}$
\caption{$(V-K)$ colours of the general population compared to the
$(V-K)$ colours of those spirals which are progenitors of
present-day early-type galaxies. Black dots represent spiral
galaxies and red dots represent spiral progenitors. The left-hand
column shows the spiral $(V-K)$ CMR from $z=0$ to $z=1$. In the
right hand column we show the fraction of spirals in parts of the
$(V-K)$ CMR space which are progenitors of early-type galaxies.
The fractions are indicated using the colour coding shown in the
legend. Warmer colours indicate a higher progenitor fraction (red
implies a progenitor fraction of 1, black represents a progenitor
fraction of 0 and parts of the CM space without any galaxies are
not colour-coded).} \label{fig:vk_spiralprogs}
\end{minipage}
\end{center}
\end{figure*}

At intermediate redshift ($z\sim0.5$), \emph{red} spirals, with
$-21.5<M(B)<-20.5$ and $(B-V)>0.6$, have $\sim30$ percent
probability of being an early-type progenitor, while blue spirals
in the same luminosity range have a low progenitor probability.
For larger spirals at these redshifts, the probabilities are
appreciably higher - red spirals with $(B-V)>0.7$ have between a
75 and 95 percent chance of being progenitors, while 50 to 75
percent of blue spirals in this luminosity range are progenitors.
The situation at high redshift $z\sim1$ is similar to that at
intermediate redshift. For completess we also show, in Figure
\ref{fig:vk_spiralprogs}, the corresponding plot for the $(V-K)$
colours for the spiral population - the trends are similar to
those we have just described in $(B-V)$.

\section{Progenitor evolution in clusters}
Before the advent of large scale surveys, dense regions of the
Universe were often targetted for early-type galaxy studies, both
at low and high redshift
\citep[e.g.][]{BLE92,Stanford1998,VD1998,VD1999,VD2000,VD01,Blakeslee2003}.
While studies of dense regions are attractive for a variety of
reasons \citep[e.g.][]{Ellis2002,VD2004}, a key benefit is
observational (statistical) convenience - clusters provide access
to large homogeneous samples of luminous objects at all redshifts.
It has been usual to `connect' results from cluster studies over
large redshift ranges to determine (at least qualitatively) the
chronology of galaxy evolution.

In this section we investigate progenitors of present-day
\emph{cluster} early-types, which are themselves in \emph{dense}
regions at $z>0$. The motivation for this investigation is two
fold. Firstly (and most importantly), it provides a comparison to
the vast literature of `cluster' early-type studies. Secondly,
this version of the GALICS model has been accurately calibrated to
match the optical CMRs of early-types in \emph{dense} regions from
low to high redshift \citep{Kaviraj2005a}, which implies that the
colours of early-type \emph{progenitors} are also reasonably well
constrained. In short, the GALICS model reproduces the optical
colours of cluster early-types at $z=0$ \emph{and} the colours of
all progenitors (regardless of morphology) \emph{in dense regions}
at high redshift. Hence, our analysis of the evolution of the
\emph{complete} progenitor set (i.e. not just spirals as was
studied in the previous section), restricted to dense regions, is
likely to be robust.

It is important here to clarify the definitions of `density' that
we use to define both `clusters' at present-day, and `dense'
regions at high redshift. As mentioned before, model `density' is
assumed to be a direct function of the mass of the DM halo in
which galaxies are embedded. At $z=0$, a DM halo mass of
$10^{14}M_\odot$ represents the lower limit for a cluster-hosting
halo. Observational studies of dense regions at high redshift are
likely to contain an assortment of cluster-type haloes of varying
occupancies. Furthermore, DM haloes themselves are evolving - on
average, the largest haloes at $z=1$ are likely to be roughly half
their size at present day \citep[e.g][]{VandenBosch2002}. To take
these two points into account, we use a \emph{variable} lower mass
limit for `cluster-hosting' haloes at $z>0$. At a given redshift
this lower limit is calculated from the average mass accretion
history \citep[][see their Figure5]{VandenBosch2002} applied to a
$10^{14}M_\odot$ halo.

\begin{figure}
\begin{center}
\includegraphics[width=3.5in]{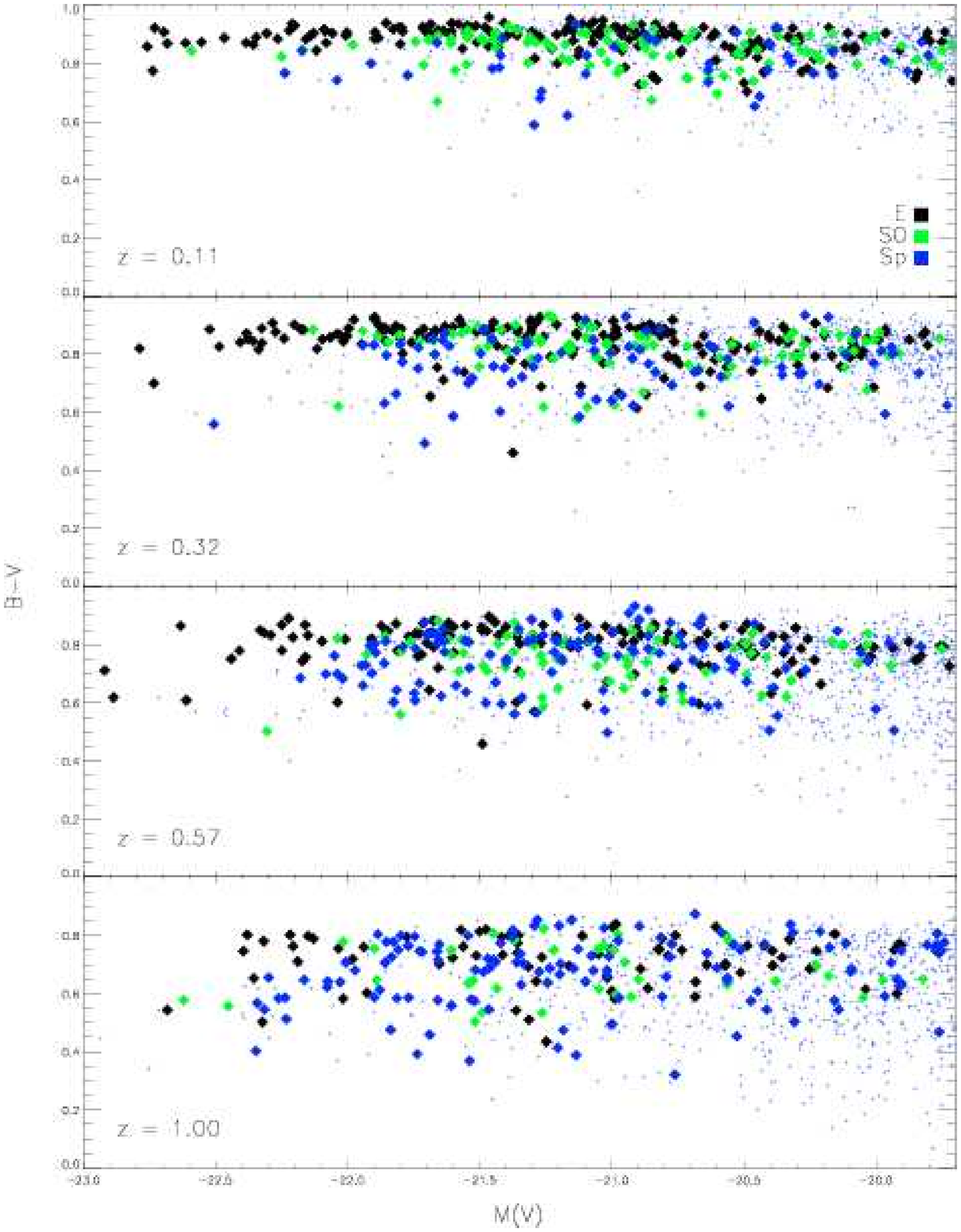}
\caption{The $(B-V)$ CMR in clusters in the redshift range
$0<z<1$. Large diamonds indicate the progenitors of present-day
cluster early-type galaxies. Small dots indicate galaxies which do
not contribute to the mass in present-day cluster early-types.
Note that all elliptical and S0 galaxies in dense regions are, not
unexpectedly, progenitors of present-day cluster early-types.}
\label{fig:cluster_cmr}
\end{center}
\end{figure}

\begin{figure}
\begin{center}
\includegraphics[width=3.5in]{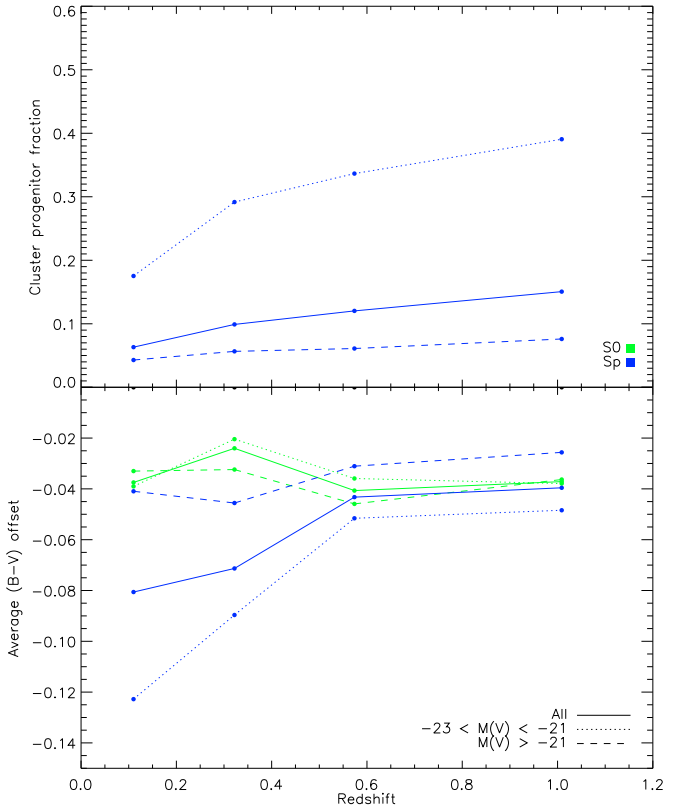}
\caption{TOP PANEL: The fraction of spiral galaxies in dense
regions at high redshift (split by luminosity) which are
progenitors of cluster early-types at $z=0$. BOTTOM PANEL: The
offset in the $(B-V)$, \emph{with respect to elliptical
progenitors}, of S0 and spiral progenitors in clusters. The
offsets are shown split by luminosity.}
\label{fig:clusterprog_props}
\end{center}
\end{figure}

Figure \ref{fig:cluster_cmr} shows the $(B-V)$ CMR in clusters in
the redshift range $0<z<1$. Large diamonds indicate progenitor
galaxies - black indicates ellipticals, green corresponds to S0s
and blue indicates spiral galaxies. Small dots indicate galaxies
which do not contribute to the mass in present-day cluster
early-types. All early-type galaxies in dense regions are, not
unexpectedly, progenitors of cluster early-types at present day.
The top panel in Figure \ref{fig:clusterprog_props} shows the
fraction of spiral galaxies in dense regions at high redshift
(split by luminosity) which are progenitors of early-types at
$z=0$. The bottom panel shows the offset in $(B-V)$, \emph{with
respect to elliptical progenitors}, of the S0 and spiral
progenitor galaxies. The offsets are shown split by luminosity.

We find that at high redshift ($z\sim1$) up to 40 percent of large
spirals ($-23<M(V)<-21$) are progenitors, whereas only $\sim$10
percent of small spirals ($M(V)>-21$) are members of the
progenitor set. Large spirals are four times more likely to be
progenitors than small spirals, regardless of redshift, in the
redshift range $0<z<1$. Elliptical galaxies form the reddest locus
in $(B-V)$. S0 galaxies show an average offset of -0.04 compared
to the elliptical population, regardless of luminosity. Large
spiral progenitors show an average $(B-V)$ offset of -0.05
compared to the elliptical population at high redshift, mainly
because the scatter in the elliptical colours also tends to be
large at high redshift. At low redshift the offset is more
pronounced - large spiral progenitors are upto 0.1 mags bluer in
$(B-V)$ than the elliptical population.


\section{The `red sequence' as a proxy for the progenitor set}
\begin{figure}
\begin{center}
\includegraphics[width=3.5in]{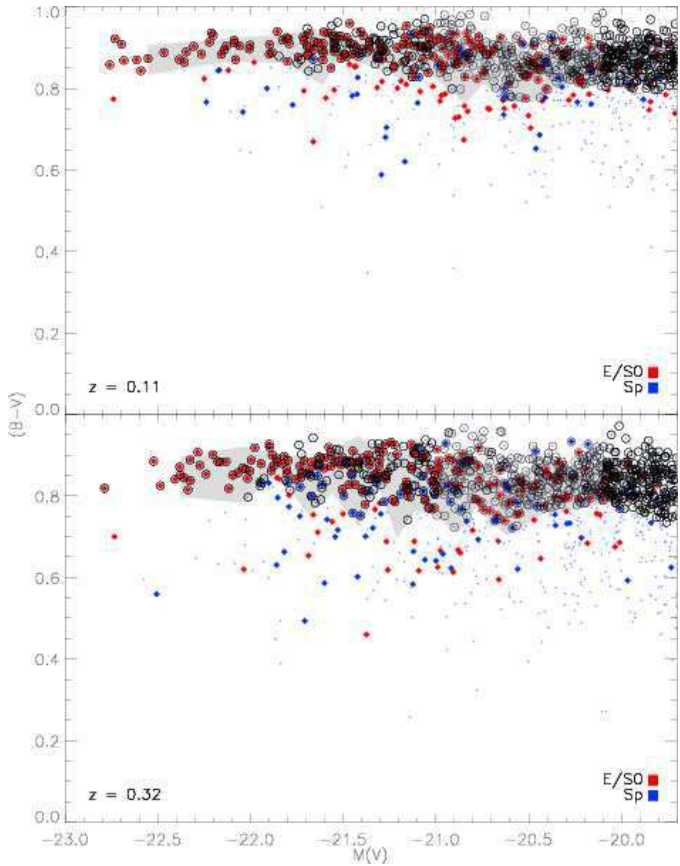}
\caption{The composition of the `red sequence', defined as the
galaxy population which occupies the part of the CM space
dominated by early-type galaxies (shown in grey), compared with
the progenitor set in cluster populations. The diamonds indicate
galaxies which are part of the progenitor set. Galaxies which are
not part of the progenitor set are shown using small crosses.
Galaxies in the red sequence are circled. The 'red sequence'
misses blue galaxies, both early-type and late-type. This plot
shows the evolution of the `red sequence' from low to intermediate
redshift.}
\label{fig:red_members1}
\end{center}
\end{figure}

\begin{figure}
\begin{center}
\includegraphics[width=3.5in]{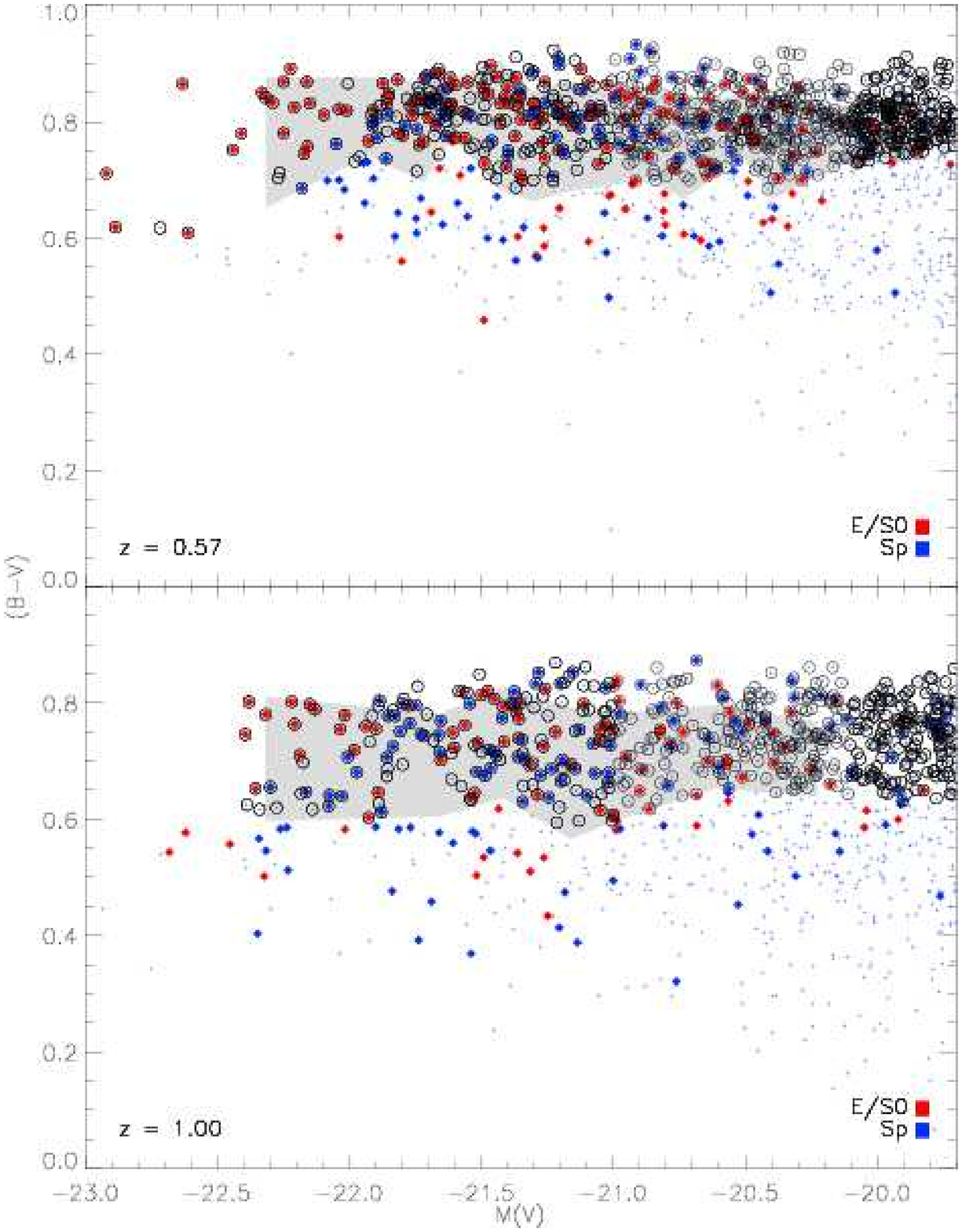}
\caption{The composition of the `red sequence', defined as the
galaxy population which occupies the part of the CM space
dominated by early-type galaxies (shown in grey), compared with
the progenitor set in cluster populations. The diamonds indicate
galaxies which are part of the progenitor set. Galaxies which are
not part of the progenitor set are shown using small crosses.
Galaxies in the red sequence are circled. The `red sequence'
misses blue galaxies, both early-type and late-type. This plot
shows the evolution of the `red sequence' from intermediate to
high redshift.}
\label{fig:red_members2}
\end{center}
\end{figure}

Early-type galaxies in clusters tend to preferentially populate
the reddest parts of the CM space. In this section we investigate
whether the sample of galaxies (without reference to morphology),
within the `red sequence' defined by the early-type population,
can be used as a proxy for the progenitor set. We define the red
sequence as the galaxy population which occupies the part of the
CM space `dominated' by early-type galaxies. In Figures
\ref{fig:red_members1} and \ref{fig:red_members2} the CM space
dominated by early-type galaxies is shown in grey - this region is
determined by a progressive one-sigma fit to the colours of the
early-type population. It therefore contains, on average, 68
percent of the early-type population within it. Large diamonds
indicate galaxies which are part of the progenitor set. Galaxies
which are not part of the progenitor set are shown using small
crosses. Galaxies in the red sequence are circled. It is apparent
that the red sequence misses blue progenitor galaxies, both
early-type and late-type. Figure \ref{fig:red_members1} shows the
evolution of the `red sequence' from low to intermediate redshift
and Figure \ref{fig:red_members2} the evolution from intermediate
to high redshift.

\begin{figure}
\begin{center}
\includegraphics[width=3.5in]{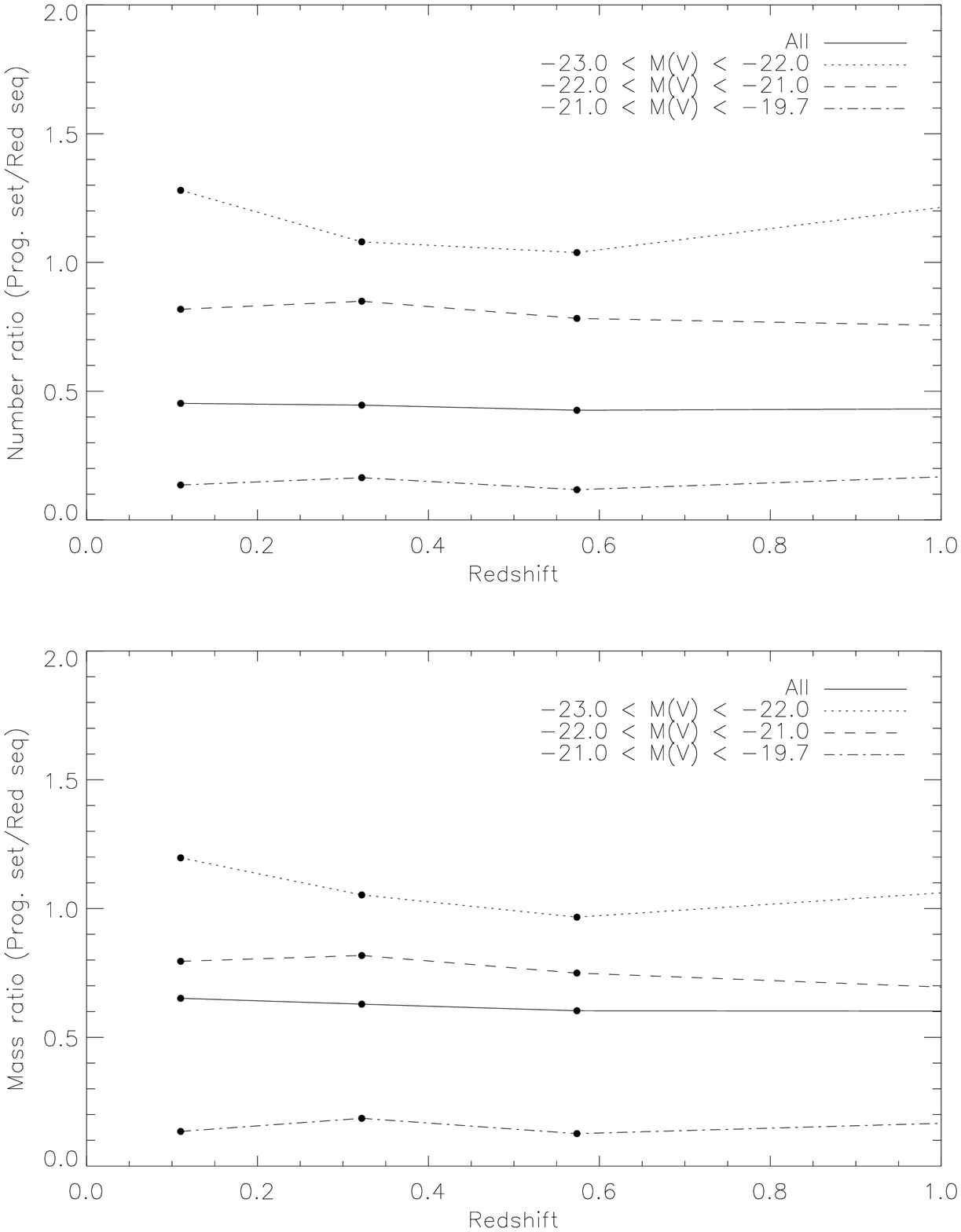}
\caption{Comparison between galaxies in the actual progenitor set
to those in the red sequence. TOP: Number ratio between the
progenitor set population and the red sequence population split by
redshift and luminosity. BOTTOM: Mass ratio between the progenitor
set population and the red sequence population split by redshift
and luminosity. It is apparent that large galaxies in the red
sequence trace the progenitor set well in terms of number and mass
but that the relationship breaks down rather rapidly as we go
towards the lower end of the luminosity function.}
\label{fig:mass_lum_ratio}
\end{center}
\end{figure}

In Figure \ref{fig:mass_lum_ratio} we compare galaxies in the
actual progenitor set to those in the red sequence. The top panel
shows the number ratio between the progenitor set population and
the red sequence population, split by redshift and luminosity,
while the bottom panel shows the mass ratio between the progenitor
set population and the red sequence population split by redshift
and luminosity. It is apparent that large galaxies
($-23<M(V)<-21$) in the red sequence trace the progenitor set well
in terms of number and mass but that the relationship breaks down
as we go towards the lower end of the luminosity function
($M(V)>-21$). The number and mass fractions remain stable, as a
function of the luminosity slices shown in Figure
\ref{fig:mass_lum_ratio}, within the redshift range explored in
this study ($0<z<1$).

Luminosity evolution studies which use the red sequence as a proxy
for the early-type population (e.g. Bell et al. 2004) can
therefore achieve accurate results \emph{only} for the upper end
of the luminosity function. However, the red sequence should not
be used as a proxy for the progenitor set further down the
luminosity function, since a larger fraction of the (spiral) red
sequence does not contribute to the progenitor set. In addition,
the red sequence, almost by definition, misses contributions due
to early-types which lie blueward of it. This is increasingly true
at higher redshift - hence conclusions based on the colours of the
red sequence should \emph{not} generally be applied to early-type
evolution.


\section{Conclusions}
We have comprehensively explored the extent of progenitor bias as
a function of the luminosity and environment of the early-type
remnant at $z=0$. Our study, using the GALICS semi-analytical
model, is the first of its kind which uses a fully realistic
semi-analytical framework, in which mass assembly and
morphological transformation can be followed accurately in the
context of the currently popular hierarchical merger paradigm.
Against the backdrop of the impending or recent release of data
from large-scale surveys \emph{at high redshift}, the results of
this study are timely, providing a picture of the mass assembly in
present-day early-types and gauging the extent by which progenitor
bias may affect the conclusions from `early-type only' studies at
high redshift.

The main conclusions from our study can be summarised as follows:

\begin{itemize}

    \item Larger early-types in all environments are \emph{assembled
    later} than their less massive counterparts. However, their
    stellar populations are generally older \citep{Kaviraj2005a}.

    \item On average, without reference to environment, only 35
    percent of early-type galaxies are in place by $z=1$.
    Morphological transformations ar significantly faster in
    cluster environments (where the vast majority of early-type
    studies have been based before the advent of large-scale
    surveys). In clusters almost 70 percent of early-types are in
    place by $z=1$. In other words, the probability of a `major
    merger', which creates an early-type remnant is low after
    $z=1$ in cluster type environments.

    \item Averaging across all environments, at $z\sim 1$, less
    than 50 percent of the stellar mass which ends up in
    early-types today is actually in early-type progenitors at
    this redshift. This value is around 65 percent in clusters
    owing to faster morphological transformation in this
    environment. In other words, looking \emph{only} at early-type
    progenitors does not take into account almost half the mass in
    the progenitor set - the progenitor set doubles in mass in the
    redshift range $0<z<1$.

    \item Progenitor bias does not arise simply because
    (late-type) progenitor mass is missed, but also because the age
    profile of mass in progenitors of different morphological
    types tend to vary. Spiral progenitors are typically `bluer'
    because they host more recently formed stars than early-type
    progenitors. Hence, age-dating the progenitor set using an
    `early-type only' CMR after excluding the spiral progenitors
    biases the CMR towards redder colours and \emph{overestimates} the
    average of the population.

    \item One of the principal aims of this study is to provide a
    means of \emph{including} spiral progenitors which might be
    early-type progenitors, and thus correct, at least partially,
    for progenitor bias. We have therefore focussed on spiral
    progenitors in the model and compared their properties, in
    detail, to the general spiral population.

    \item There is a greater preponderance of progenitors among
    larger spirals at all redshifts. At
    low redshifts ($z<0.1$), 20 to 40 percent of spirals with
    $M(B)<-20.5$ are early-type progenitors. At intermediate redshifts
    $0.3<z<0.5$, these values rise to 30 and 60 percent respectively.
    At high redshift ($z \sim 1$) spirals with $M(B)<-21.5$ have more
    than a 60 percent probability of being an early-type progenitor
    while spirals with $-20<M(B)<-21.5$ have between a 30 and 40
    percent chance of being early-type progenitors. The falling
    progenitor fractions towards lower redshift are partly due to the
    changing morphological mix of the Universe.

    \item The colour-magnitude space of the spiral population
    provides a better route to identifying spiral progenitors using both the
    luminosity and the optical colour. At $z\sim0.1$,
    spirals with $-21.5<M(B)<-20.5$ have $\sim 30$ percent chance of
    being a progenitor. For larger spirals, those with red $(B-V)$
    colours, i.e. $(B-V)>0.8$, have $\sim60$ percent chance of being a
    progenitor, while the corresponding probability for bluer spirals
    is 30 to 50 percent.

    At intermediate redshift ($z\sim0.5$), \emph{red} spirals, with
    $-21.5<M(B)<-20.5$ and $(B-V)>0.6$, have $\sim30$ percent
    probability of being an early-type progenitor, while blue spirals
    in the same luminosity range have a low progenitor probability.
    For larger spirals at these redshifts, the probabilities are
    appreciably higher - red spirals with $(B-V)>0.7$ have between a
    75 and 95 percent chance of being progenitors, while 50 to 75
    percent of blue spirals in this luminosity range are progenitors.
    The situation at high redshift $z\sim1$ is similar to that at
    intermediate redshift. The trends in the $(V-K)$ colour are similar to
    those in $(B-V)$.

    \item Finally we have explored the correspondence between the
    progenitor set and the `red sequence', defined as the part of
    the CM parameter space which is dominated by early-type galaxies.
    We find that galaxies, both late and early-type, that fall in
    this parameter space do not necessarily trace the progenitor
    set well. Large galaxies ($-23<M(V)<-21$) in the red sequence
    correspond to the progenitor set reasonably well in terms of number and mass but the
    relationship breaks down as we go towards the lower end of the luminosity function
    ($M(V)>-21$). Hence, luminosity evolution studies which use the red sequence as a proxy
    for the early-type population, therefore
    achieve accurate results only for the upper end of the luminosity
    function. In addition, the red sequence,
    almost by definition, misses contributions due to early-types which lie blueward
    of it - hence conclusions based on the colours of the red sequence should not
    generally be applied to early-type evolution, \emph{especially at
    high redshift}.

\end{itemize}


\nocite{Martin2005}
\nocite{Bell2004}


\bibliographystyle{chicago}
\bibliography{references2}


\end{document}